\documentclass[preprint]{aastex63}
\usepackage{etoolbox}
\usepackage{graphicx}
\usepackage{natbib}

\newcommand{\gsim}{\mbox{\hspace{.2em}\raisebox{.5ex}{$>$}\hspace{-.8em}\raisebox{-.5ex}{$\sim$}\hspace{.2em}}}
\newcommand{\lsim}{\mbox{\hspace{.2em}\raisebox{.5ex}{$<$}\hspace{-.8em}\raisebox{-.5ex}{$\sim$}\hspace{.2em}}}

\newcommand{\E}[1]{\times 10^{#1}}
\newcommand{\twCO}{$^{12}$CO}  \newcommand{\thCO}{$^{13}$CO}

\newcommand{\HI}{\mbox{H\,\textsc{i}}}

      \newcommand{\ps}{\,{\rm s}^{-1}}
    \newcommand{\Msun}{M_{\odot}}   
    \newcommand{\km}{\,{\rm km}}
\newcommand{\kpc}{\,{\rm kpc}}

\begin{document}

\title{
Revealing Gas Inflows Towards the Galactic Central Molecular Zone
}

\shorttitle{Large-scale Gas Inflows towards the CMZ}

\correspondingauthor{Yang Su}
\email{yangsu@pmo.ac.cn}

\author[0000-0002-0197-470X]{Yang Su}
\affil{Purple Mountain Observatory and Key Laboratory of Radio Astronomy,
Chinese Academy of Sciences, 10 Yuanhua Road, Nanjing 210023, China}
\affiliation{School of Astronomy and Space Science, University of Science and
Technology of China, 96 Jinzhai Road, Hefei 230026, China}

\author{Shiyu Zhang}
\affiliation{Purple Mountain Observatory and Key Laboratory of Radio Astronomy,
Chinese Academy of Sciences, 10 Yuanhua Road, Nanjing 210023, China}
\affiliation{School of Astronomy and Space Science, University of Science and
Technology of China, 96 Jinzhai Road, Hefei 230026, China}

\author{Yan Sun}
\affiliation{Purple Mountain Observatory and Key Laboratory of Radio Astronomy,
Chinese Academy of Sciences, 10 Yuanhua Road, Nanjing 210023, China}
\affiliation{School of Astronomy and Space Science, University of Science and
Technology of China, 96 Jinzhai Road, Hefei 230026, China}

\author{Ji Yang}
\affiliation{Purple Mountain Observatory and Key Laboratory of Radio Astronomy,
Chinese Academy of Sciences, 10 Yuanhua Road, Nanjing 210023, China}
\affiliation{School of Astronomy and Space Science, University of Science and
Technology of China, 96 Jinzhai Road, Hefei 230026, China}

\author{Qing-Zeng Yan}
\affiliation{Purple Mountain Observatory and Key Laboratory of Radio Astronomy,
Chinese Academy of Sciences, 10 Yuanhua Road, Nanjing 210023, China}

\author{Shaobo Zhang}
\affiliation{Purple Mountain Observatory and Key Laboratory of Radio Astronomy,
Chinese Academy of Sciences, 10 Yuanhua Road, Nanjing 210023, China}

\author{Zhiwei Chen}
\affiliation{Purple Mountain Observatory and Key Laboratory of Radio Astronomy,
Chinese Academy of Sciences, 10 Yuanhua Road, Nanjing 210023, China}

\author{Xuepeng Chen}
\affiliation{Purple Mountain Observatory and Key Laboratory of Radio Astronomy,
Chinese Academy of Sciences, 10 Yuanhua Road, Nanjing 210023, China}
\affiliation{School of Astronomy and Space Science, University of Science and
Technology of China, 96 Jinzhai Road, Hefei 230026, China}

\author{Xin Zhou}
\affiliation{Purple Mountain Observatory and Key Laboratory of Radio Astronomy,
Chinese Academy of Sciences, 10 Yuanhua Road, Nanjing 210023, China}

\author{Lixia Yuan}
\affiliation{Purple Mountain Observatory and Key Laboratory of Radio Astronomy,
Chinese Academy of Sciences, 10 Yuanhua Road, Nanjing 210023, China}

\begin{abstract}
We study the gas inflows towards the Galactic Central Molecular Zone (CMZ) 
based on the gas morphological and kinematic features from the MWISP 
in the region of $l=1\fdg2-19\fdg0$ and $|b|\lsim3\fdg0$.
We find that the near dust lane appears to extend to $l\sim15^{\circ}$, 
in which the end of the large-scale gas structure 
intersects with the 3~kpc-ring at a distance of $\sim 5$~kpc.
Intriguingly, many filamentary molecular clouds (MCs), 
together with the bow-like/ballistic-like
clouds and continuous CO features with notable velocity gradient, 
are finely outlined along the long structure.
These MCs also have relatively large velocity dispersions, indicating 
the shocked gas generated by local continuous accretion and
thus the enhanced turbulence along the entire gas structure.
We suggest that the $\sim$3.1--3.6~kpc-long CO structure originates from 
the accretion molecular gas driven by the Galactic bar.
The gas near the bar end at the 3~kpc-ring region becomes
an important reservoir for the large-scale accreting flows inwards to
the CMZ through the bar channel.
The inclination angle of the bar is estimated to be 
$\phi_{\rm bar}=23^{\circ}\pm3^{\circ}$, while the pattern speed of the bar is 
$\Omega_{\rm bar}\lsim32.5\pm2.5\ \km\ps\kpc^{-1}$. The total mass of the whole 
near gas lane is about 1.3$\pm0.4\times10^{7}\ \Msun$ according to the calculated 
$X_{\rm CO}\sim 1.0\pm0.4\E{20}$~cm$^{-2}$(K~km~s$^{-1})^{-1}$ from the large-scale 
\twCO\ and \thCO\ data and the complementary \HI\ data.  
We revisit the gas inflow rate as a mean value of 1.1$\pm0.3\ \Msun$~yr$^{-1}$,
which seems to be comparable to the outflow's rate of the Galactic
nuclear winds after applying the updated lower X-factor above.
\end{abstract}


\section{Introduction}
Galaxies are giant and delicate ecosystems in which material flows in and out. 
As the closest galaxy to us, the Milky Way becomes a unique laboratory for
exploring the physical details of the baryon cycle. 
Benefiting from large-scale surveys in multiwavelength, many exciting discoveries 
have been made in observational and theoretical studies.
One of the most important findings is the revelation of the existence of
the Fermi bubbles \citep[FBs; i.e.,][]{2010ApJ...724.1044S} that are directly
related to the Galactic nuclear winds \citep[e.g.,][]
{1984Natur.310..568S,2003ApJ...582..246B,2010ApJ...717..825D,
2019Natur.573..235H,2019Natur.567..347P,2021A&A...646A..66P,
2020Natur.588..227P,2024A&ARv..32....1S}. 
The FBs, together with the combined features in multiwavelengths and
multiple spatial scales, are believed to be associated with past energetic events
(e.g., nuclear activities and intense processes of massive star formation)
near the Galactic Center (GC).

The cold atomic and molecular gas related to the Galactic nuclear winds has 
attracted even more attention very recently due to the cold gaseous substances 
found in the FBs from large-scale \HI\ and CO studies
\citep[e.g.,][]{2016ApJ...826..215L,2018ApJ...855...33D,2020ApJ...888...51L,
2020Natur.584..364D,2023MNRAS.524.1258N,2023A&A...674L..15V}.
Using the CO data, we also find that the Galactic nuclear winds
have a significant impact on the Galactic gaseous disk,
leading to material accumulation at the 3-kpc crater-walls 
\citep[][]{2022ApJ...930..112S} 
and gas deficiency in the region of 0.3~kpc$\lsim R_{\rm GC}\lsim$~3.1~kpc
\citep[here $R_{\rm GC}$ is the Galactocentric distance; e.g.,][]
{1976ARA&A..14..275B,1990ARA&A..28..215D,2001ApJ...547..792D,
2009ARA&A..47...27K,2021ApJ...910..131S}.
The mass-loading rate of the outflows related to the Galactic nuclear winds 
is estimated to be $\sim2-4\ \Msun$~yr$^{-1}$ \citep[][]{2022ApJ...930..112S}.
Considering additional warm and hot gas in the FBs 
\citep[][]{2016ApJ...829....9M,2020ApJ...899L..11K,2020Natur.588..227P,2021ApJ...915...85Z},
the actual mass-loading rate of the Galactic nuclear winds may be  
even higher than what is inferred from the cold atomic and molecular gas alone.

On the other hand, there is substantial molecular gas hosting in the region 
of $R_{\rm GC}\lsim$~300~pc (i.e., the Central Molecular Zone (CMZ)). 
The total molecular gas in the CMZ is estimated to be 2--6$\times10^{7}\ \Msun$
\citep[e.g., see][]{1998A&A...331..959D,2007A&A...467..611F}, 
corresponding to 3\%--10\% of all molecular gas in the Galaxy.
The origin of the CMZ is probably related to the episodic accumulation
of the gas inflows driven by the gravitational field of the Galactic rotating bar 
\citep[see the comprehensive review in][]{2023ASPC..534...83H}.
That is, the molecular gas supply along the dust lane accretion (or the 
bar channel) may account for the large amount of material accumulation in the CMZ.
The gas inflow rate onto the CMZ is estimated to be about 0.4--2.7~$\Msun$~yr$^{-1}$
at a dynamical timescale of tens of Myr \citep[][]{2012MNRAS.423.3512C,2019MNRAS.490.4401A,
2019MNRAS.484.1213S,2020MNRAS.499.4455T,2021ApJ...922...79H}.

The above studies show that there seems to be an overall balance 
between mass inflow and outflow for the inner 3~kpc region 
of the Milky Way in the current period if we ignore the low star formation rate 
of $\sim$0.1~$\Msun$~yr$^{-1}$ \citep[][]{2021NewAR..9301630B} in the CMZ.
However, the estimations of the mass inflow and mass outflow rates are 
model dependent and are somewhat uncertain.
Therefore, an in-depth and full study of the physical processes of 
the accretion gas flowing inward to the CMZ is important for
understanding the gas cycle in the Milky Way.
In particular, the comparisons between spatial and kinematic features 
in observations and dedicated simulations are fundamental and crucial 
in revealing new details and constraining model parameters.

In the Letter, we focus on the large-scale inflows of molecular gas 
towards the CMZ based on the $l$-$b$-$v$ CO data from the 
Milky Way Imaging Scroll Painting (MWISP) survey.
We describe the CO and \HI\ data in Section 2.
In Section 3, we investigate the distribution and kinematics of molecular gas 
along the near dust lane on a high dynamic range and at $\sim$~1--2~pc 
physical resolution \citep[i.e., adopting $R_{0}$=8.15~kpc from][]{Reid19}, 
which is about 2 orders of magnitude better than that of the GMCs
in nearby galaxies \citep[e.g.,][]{2024arXiv240319843S}.
Our new data clearly show that a large amount of gas is flowing inwards
and bringing the gas to the CMZ. 
We construct a toy model to quantitatively delineate the gas kinematics 
and give a schematic diagram to explain the CO observations.
We discuss the connection between the gas inflow driven by 
the Galactic bar and outflow from the Galactic nuclear winds in a whole picture. 
Finally, we summarize the new findings and results in Section 4.

\section{CO and \HI\ data}
The CO data set studied here is from the MWISP 
project \citep[i.e., see details in][]{2019ApJS..240....9S}.
In short, the MWISP CO data are observed in a region of 
$l=1\fdg2-19\fdg0$ and $|b|\lsim3\fdg0$, which fully covers the molecular gas 
in the near dust lane with spatial and spectral resolutions of $\sim50''$ 
and $\sim$~0.2~$\km\ps$, respectively. After cleaning the data, the reduced 
3D cubes (i.e., the position-position-velocity space (PPV)) with a 
grid spacing of 30$''$ have a typical root mean square (rms) noise level of 
$\lsim$~0.5~K for \twCO\ and $\lsim$~0.3~K for \thCO/C$^{18}$O, respectively. 

In this work, the \HI\ data from the all-sky \HI\ survey 
\citep[][]{2005A&A...440..775K,2016A&A...594A.116H} are used for 
large-scale comparisons with the MWISP CO data.
The angular and velocity resolutions of the \HI\ data are 
16\farcm2 and 1.5~$\km\ps$, respectively. The typical RMS sensitivity
of the HI4PI data is $\sim$43~mK.

\section{Results and Discussions}
\subsection{Molecular Gas Flowing towards the CMZ along the Near Dust Lane}
Thanks to many spectroscopic data from large surveys 
\citep[e..g,][]{2001ApJ...547..792D,2005A&A...440..775K,2010A&A...523A..45R,
2012MNRAS.419.2961J,2016A&A...594A.116H,2016ApJ...831..124M,2019PASJ...71S..19T,
2020MNRAS.498.5936E,2020ApJS..248...24G,2021MNRAS.500.3064S} and many excellent 
research efforts on the Milky Way, significant progress has been made in the study 
of the properties and motion of gas in and around the Galactic CMZ in the last two decades
 \citep[e.g.,][]{2006A&A...455..963R,2006A&A...447..533L,2008A&A...486..467L,
2007A&A...467..611F,2008Dame,2008A&A...489..115R,2010A&A...523A..51R,2011ApJ...735L..33M,
2015MNRAS.446.4186S,2016MNRAS.457.2675H,2018A&A...613A..42R,2019MNRAS.488.4663S,
2021MNRAS.502.5896A,2022A&A...668A.183B,2023ApJ...959...93G}.
On the other hand, a lot of numerical simulations have done a good job of describing 
the distribution and evolution of gas in the CMZ and even the Milky Way 
\citep[e.g.,][]{1999A&A...345..787F,2016ApJ...824...13L,2022ApJ...925...71L,
2015MNRAS.449.2421S,2018MNRAS.475.2383S,2020MNRAS.499.4455T,2021ApJ...922...79H}.

Combining observations and theories, we know that the gravitational potential 
plays a dominant role in determining the motion of the stars and gas in our Galaxy 
\citep[see, e.g.,][]{1999MNRAS.304..512E,2015ApJ...814...13M,2016ARA&A..54..529B}. 
Towards the $R_{\rm {GC}}\lsim$3~kpc region, the Galactic bar with a 
nonaxisymmetric structure is particularly important to transport matter from 
the Galactic disk to the CMZ and then to the GC
\citep[see the recent reviews of][]{2016ARA&A..54..529B,2023ASPC..534...83H,
2024arXiv240319843S}.

However, it is difficult to carefully characterize the structure, distribution, 
and motion of the gas flows towards the CMZ by observation because of
our embedded view through the whole Galactic plane.
The MWISP CO survey, which has large-scale and high-sensitivity
triple CO isotope line data, provides an excellent opportunity for us to delve 
into the distribution and kinematics of the molecular gas flowing towards the CMZ.

\begin{figure}
\gridline{
  \hspace{-7.5ex} \fig{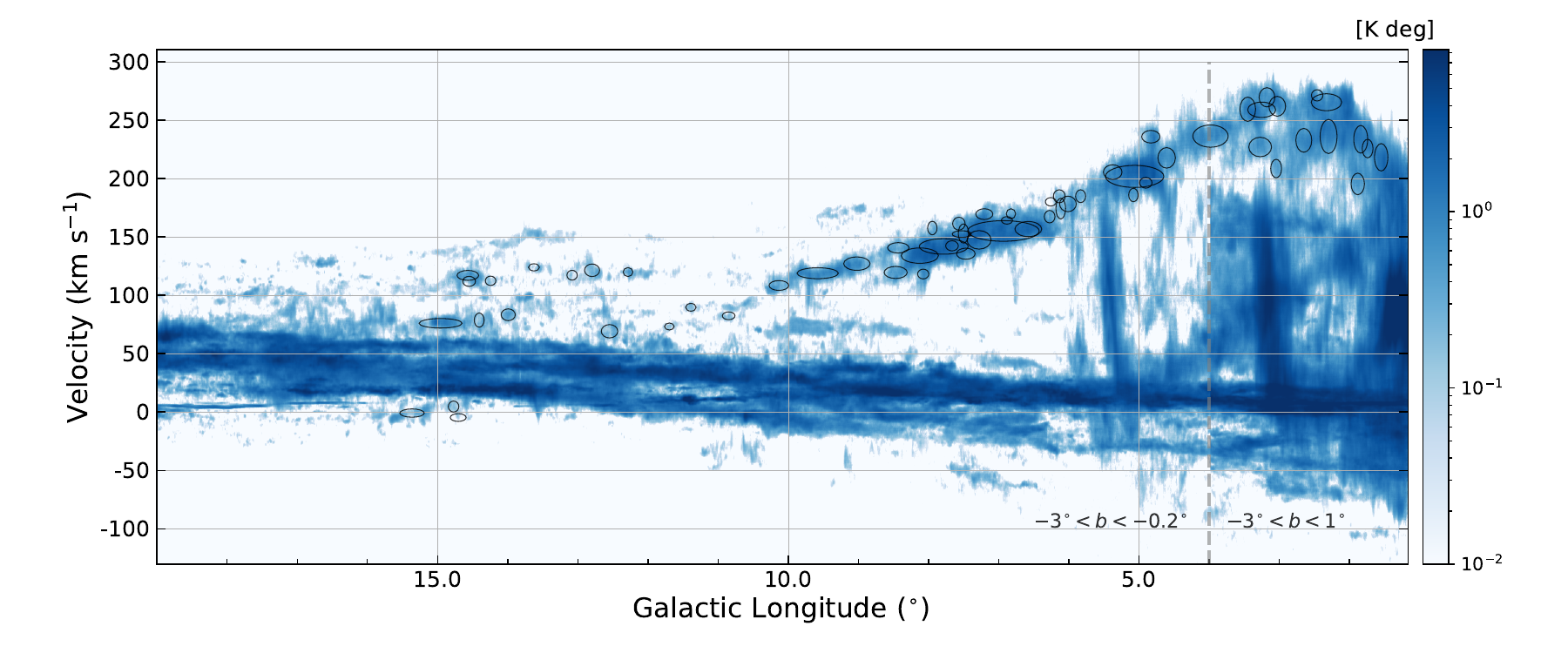}{1.1\textwidth}{} 
         }
\vspace{-13ex}
\gridline{
  \hspace{-4.5ex} \fig{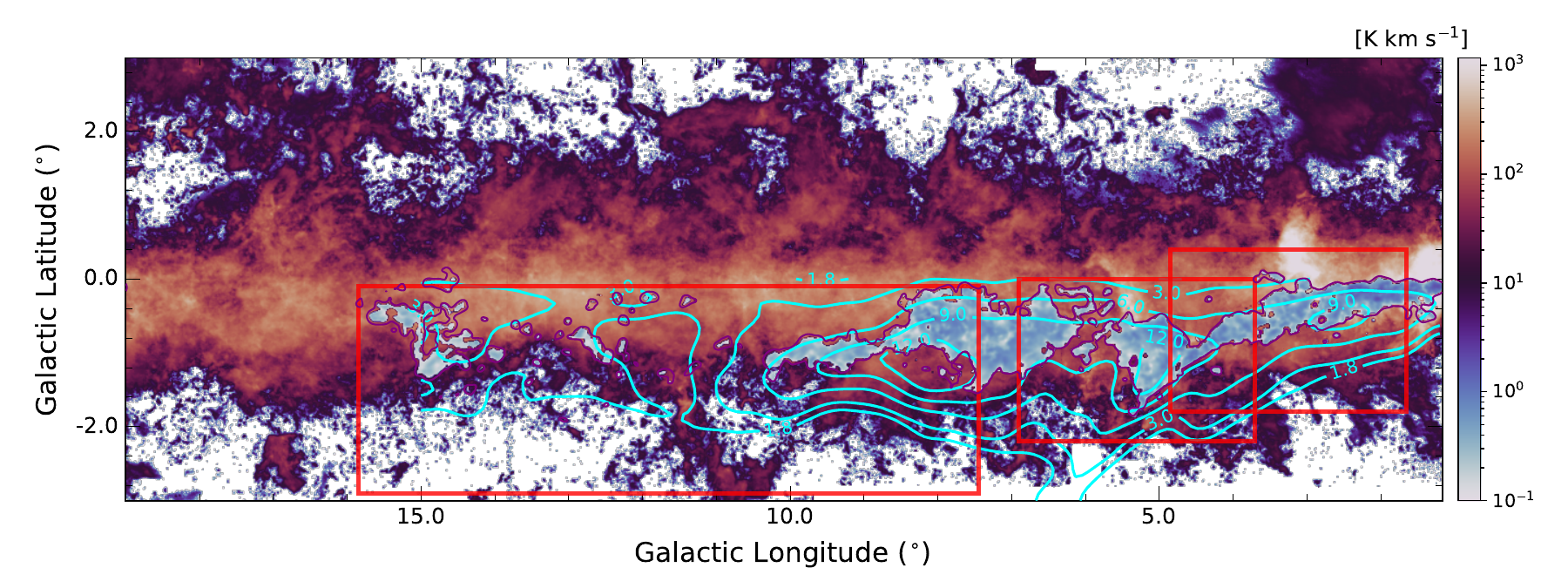}{1.045\textwidth}{} 
         }
\vspace{-7ex}
\caption{
The upper panel shows the $l$-$v$ diagram of MWISP \twCO\ emission in the region 
[$l=1\fdg2$ to $4\fdg0$, $b\leq1\fdg0$] and [$l=4\fdg0$ to $19\fdg0$, $b\leq-0\fdg2$],
while the bottom panel displays the integrated \twCO\ emission of the entire
$l$-$b$-$v$ region, overlaid with the gas associated with the
whole near dust lane (see Section 3.1).
Note that the gas inflows of the large-scale gas structure towards the CMZ 
is highlighted by the cyan part for the CO emission from the dust lane
(i.e., $I(\rm {CO})/100$)
and the cyan contours for the corresponding \HI\ emission 
scaled by $I({\rm {\HI}})\times{\rm {tan}}(b)$.
The black ellipses in the upper panel represent the 
64 large MC structures (i.e., angular size $>50\ {\rm {arcmin}}^2$, 
and 2$R_{\rm {MC}}\times v_{\rm {FWHM}}$ for the ellipse size;
here $R_{\rm {MC}}$ is the effective radius and $v_{\rm {FWHM}}$ 
the mean full width at half-maximum)
related to the whole near dust lane, and the red rectangles 
in the bottom panel represent the interesting regions shown in Figure~\ref{fig:f2}.  
\label{fig:f1}}
\end{figure}

Figure~\ref{fig:f1} shows the \twCO\ $l$-$v$ map in the range of
[$l= 1\fdg2$ to $4\fdg0$, $-3\fdg0 \lsim b \lsim +1\fdg0$]
and [$l= 4\fdg0$ to $19\fdg0$, $-3\fdg0 \lsim b \lsim -0\fdg2$] 
and the total integrated emission from the MWISP data.
The elongated gas structure, which is sketched by the cyan color 
for the CO emission and the cyan contours for the corresponding \HI\ 
emission in the bottom panel of Figure~\ref{fig:f1}, is perfectly 
delineated along the near dust lane in $l\sim[1\fdg2, 10\fdg3]$ 
and $v_{\rm {LSR}}\sim[100\km\ps, 280\km\ps]$.
Indeed, the dust lanes (or the connecting arms) are believed to directly
associate with the Galactic bar \citep[][]{1999A&A...345..787F,2006A&A...447..533L,
2008A&A...486..467L,2006A&A...455..963R,2008A&A...477L..21M,2019MNRAS.484.1213S}.

The major part of the large-scale gas structure lies slightly below 
the Galactic plane with a mean value of $b\approx -0\fdg7$ 
(or $\sim-70$~pc below the plane; see Section 3.2.2).
This striking feature, together with the considerable velocity dispersion and 
velocity gradient of these coherent clouds, is why we are able to figure this
large-scale gas structure out of the complicated CO emission 
in the inner Galaxy relatively easily.

\begin{figure}
\vspace{-1ex}
\gridline{\hspace{-8ex}  \fig{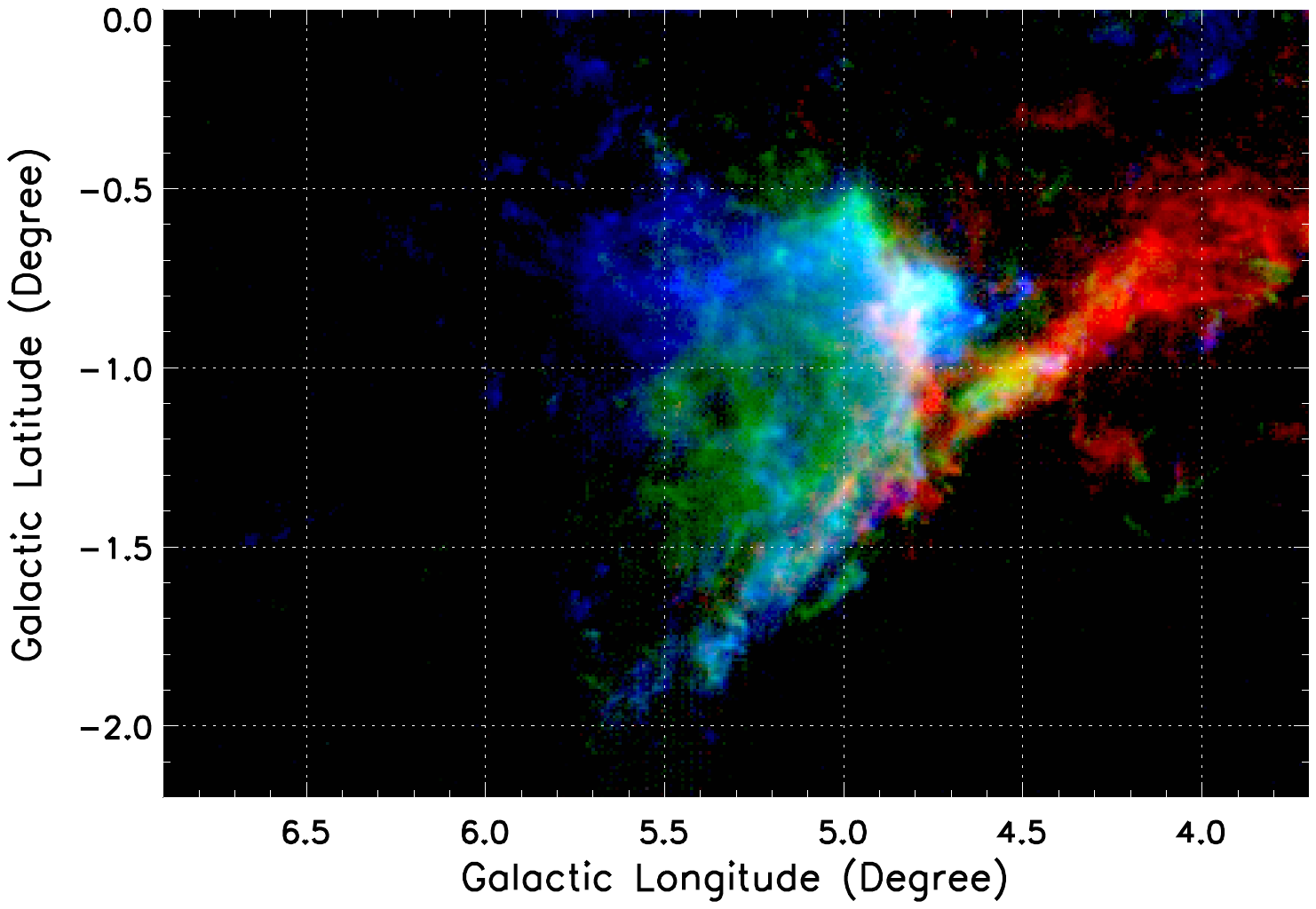}{0.6\textwidth}
           {\vspace{-52.0ex} \hspace{+11ex}  (a) RGB map for bow-like and arc-like MCs.} 
         \hspace{-13ex}  \fig{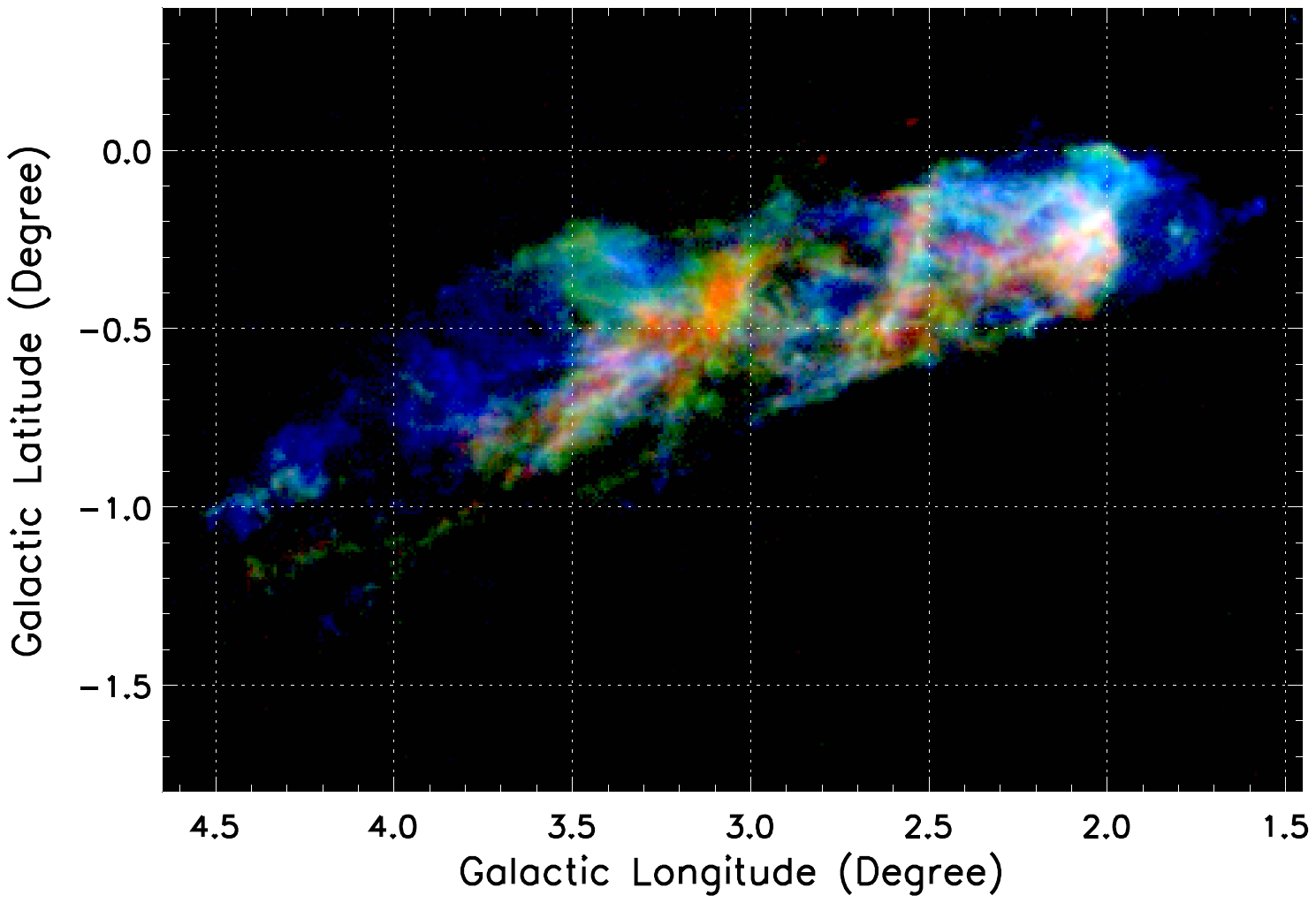}{0.6\textwidth}
           {\vspace{-52.0ex} \hspace{+9ex}  (b) RGB map for long ballistic-like MCs.}
         }
\vspace{-30.0ex}
\gridline{\hspace{-8ex} \fig{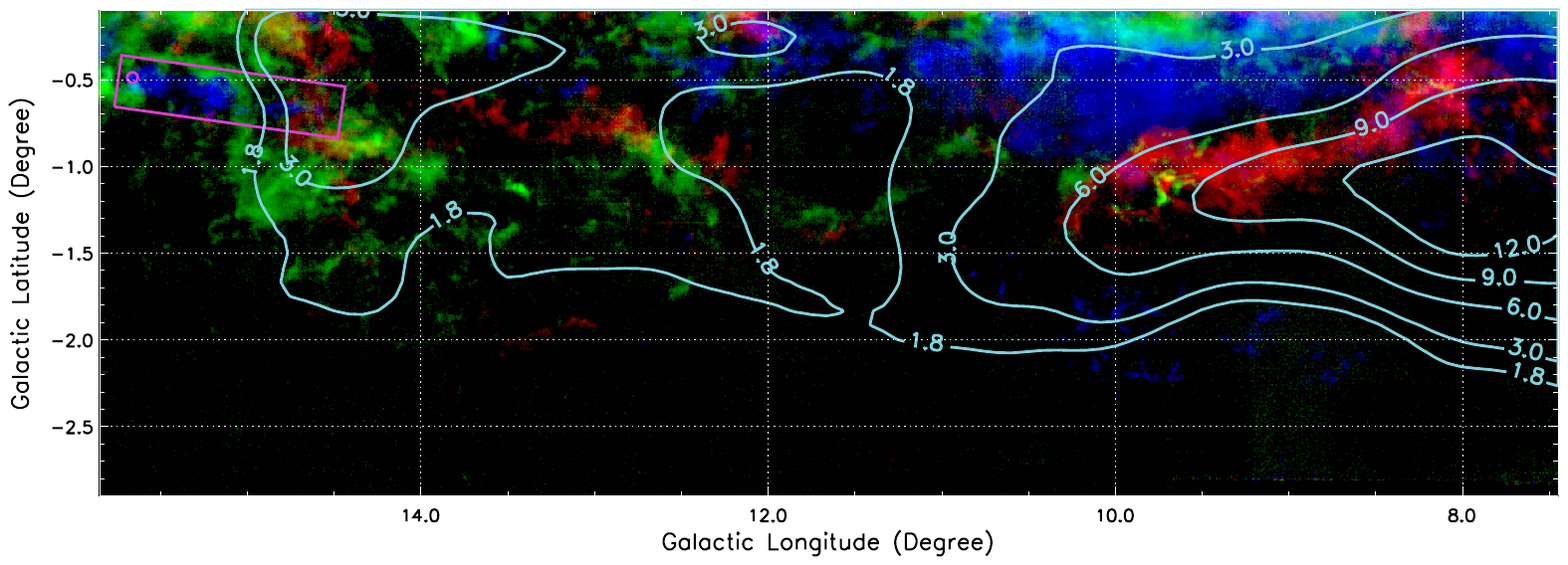}{1.14\textwidth}
{\vspace{-75.5ex} \hspace{+5ex} (c) RGB map for MCs near the end of the large-scale CO inflows.}
         }
\vspace{-29ex}
\caption{
Integrated \twCO\ emission of three interesting MC structures 
of the large-scale CO inflows towards the CMZ.
Panel (a) shows the arc-like MCs with a sharp edge towards the CMZ
in the velocity interval of [192, 202] $\km\ps$ (blue), [203, 216] $\km\ps$ (green), 
and [222, 237] $\km\ps$ (red). 
Panel (b) shows the long ballistic-like MCs with the bright head towards the CMZ
in the velocity interval of [245, 255] $\km\ps$ (blue), [256, 264] $\km\ps$ (green),
and [265, 290] $\km\ps$ (red).
Panel (c) shows the MCs located near the end of the CO flows
in the velocity interval of [$-31$, 7] $\km\ps$ (blue), [65, 105] $\km\ps$ (green),
and [110, 130] $\km\ps$ (red). The purple circle and the rectangle represent 
the maser G015.66$-$00.49 and the $\sim 0-7\km\ps$ filamentary MCs 
located at the bar end, respectively. 
The cyan contours are the same as those in Figure~\ref{fig:f1} for the 
corresponding \HI\ emission related to the WGL. 
Note that the blue color in panel 
(c) roughly traces the near-3~kpc-ring emission \citep[see][]{2008Dame}
or the lateral arm indicated in Figure~\ref{fig:cmz}.
\label{fig:f2}}
\end{figure}

Remarkably, the MWISP data clearly reveal many gas structures 
associated with the near dust lane--for example, the arc-like and bow-like 
MCs (Figure~\ref{fig:f2}a), as well as the ballistic-like CO structure 
with a length of $\gsim$800~pc (Figure~\ref{fig:f2}b). 
The moving gas scenario towards the Galactic CMZ is further supported 
by the detection of the shocked gas structures with a sharp edge towards 
the GC and the increasing velocity and certain velocity gradient along the dust lane.

That is, the gas is moving away from the observer and is approaching the CMZ
with considerable acceleration along the Galactic bar, leading to
special CO morphologies (e.g., arc-like, bow-like, and ballistic structures
in Figure~\ref{fig:f2}) and prominent kinematic features (e.g., the very wide 
CO line broadening or the large velocity dispersion and the continuous gas 
distribution with a certain velocity gradient shown in Figure~\ref{fig:f1}).
These results provide an excellent indication that the molecular gas flow 
is heading towards the CMZ region in the Galactic bar potential.
Based on the filamentary CO structures towards the intersection 
region (i.e., Sgr~E) between the far dust lane and the CMZ, 
\cite{2022ApJ...939...58W} recently proposed a similar scenario that the gas 
is being stretched and accelerated by the bar potential while falling towards the CMZ.

In addition to the above evidence of the gas flows towards the CMZ,
some signs show that the large-scale CO structure has extended to 
$l\sim 15^{\circ}$ in the end of the CO lane and is probably accreting 
the gas near the 3~kpc-ring \citep[i.e., refer to the 3~kpc arm in][]{2008Dame}.
We have compiled the observational evidence below. \\
 
1. Several CO structures at $l\gsim 10^{\circ}$ are found to be connected 
in the $l$-$v$ map, indicating the presence of a large-scale spatially continuous 
gas structure associated with the known near dust lane with a similar velocity gradient. 
These CO structures often display prominent filamentary morphology along the Galactic 
longitude at negative latitude (Figure~\ref{fig:f2}c), 
which is exactly the same as the distribution of 
the known near dust lane in $l\sim 1\fdg2$ to $10\fdg3$.  
These CO structures, together with many cometary-like and arc-shaped features, 
have a relatively large velocity dispersion, showing the enhanced turbulence 
(or surrounding shock process) in the whole region.  \\  

2. In fact, some MCs with relatively large CO line dispersion exhibit long-shell and
large arc structures in $l\sim[11^{\circ}, 15^{\circ}], b\sim[-1\fdg5, -0\fdg3],
v_{\rm {LSR}}\sim[70 \km\ps, 130 \km\ps]$, as well as many bullet CO features 
embedded in these structures
(see green and red colors for CO emission and the cyan contours for \HI\ emission 
in Figure~\ref{fig:f2}c).
These $l$-$b$-$v$ features can be interpreted as the bar-spiral intersection 
at the low gravitational potential of the bar end,
leading to the accumulation of a large amount of trapped gas with different 
velocity in the low-shear region.   \\

3. We find that the large-scale CO structure appears to converge with the near 
3~kpc-ring at $l\sim 15^{\circ}$ and $v_{\rm {LSR}}$ of tens of kilometers per second, 
probably indicating a connection between them.
Remarkably, a maser G015.66-00.49 \citep[at a parallax of 0.22$\pm0.029$~mas 
and $v_{\rm {LSR}}=-4\km\ps$,][]{Reid19} is found to be associated with the  
elongated MCs (see the purple rectangle in Figure~\ref{fig:f2}c) near the end of 
the large-scale CO structure, suggesting a distance of $\sim 5$~kpc for
the bar-spiral intersection region.  \\  

According to the above points, we suggest that the large-scale CO structure 
(or the whole gas lane (WGL)) extends to $l\sim 15^{\circ}$ and $b\sim -0\fdg5$,
connecting and accreting gas in the vicinity of the 3~kpc-ring at a distance of $\sim 5$~kpc, 
and then transporting material inward along the bar channel.
Meanwhile, in the region where the gravitational potential of the bar is large,
the moving gas is dominated by the Galactic bar to form the strong compression 
shock (Figure~\ref{fig:f2}a) and even the bound radial gas flow 
(Figure~\ref{fig:f2}b) into the CMZ.
The observational evidence related to the moving gas is also manifested in 
the prevalent shock structures with the bright head towards the CMZ
(see panels a and b in Figure~\ref{fig:f2}),
as well as the velocity gradient along the WGL revealed
by the MWISP large-scale CO observation (see Figure~\ref{fig:f1}).
Based on high-quality MWISP data, we have fully revealed these new 
observational details for the first time, supporting the physical scenario 
of bar-driven gas inflows from the 3~kpc-ring to the CMZ.

We identified more than 800 MC structures with an angular size
larger than 2.25~arcmin$^2$ (9 pixels) related to the WGL
based on Gaussian decomposition and clustering algorithm
\citep[see details in][]{2019MNRAS.485.2457H,2019A&A...628A..78R,2024AJ....167..220Z}.
Here, the smoothing parameters of $\alpha$1=1.8 and $\alpha$2=5.4 
were used for the Gaussian fitting for the \twCO\ data with the 3D grid 
of $0.5'\times0.5'\times0.5\km\ps$.
We have marked 64 large MC structures (angular size $>50\ {\rm {arcmin}}^2$) 
on the upper panel of Figure~\ref{fig:f1}. The typical excitation temperature 
of the gas is $\gsim8\pm1$~K (assuming the beam filling factor is $\lsim$1 
for these distant MCs at about 5--8~kpc) 
according to the peak emission of the 64 large \twCO\ structures. 
More than 70\% of samples in the 64 MCs have a relatively small 
\thCO-to-\twCO\ intensity ratio (i.e., $I(^{13}$CO)/$I(^{12}$CO)$\sim$0.1), 
which is roughly half the value of the normal MCs in the Galactic plane 
\citep[$\sim$0.2 in previous studies; see e.g.,][]{2019PASJ...71S..19T,2023AJ....166..121W}.
We also find that the closer the MCs in the WGL are to the GC, 
the greater the velocity dispersion they have. A detailed comparison of 
the MCs' properties within the $R_{\rm {GC}}\lsim3$~kpc region will be shown elsewhere.

\subsection{Parameters of Moving Gas towards the CMZ}
\begin{figure}
\gridline{\fig{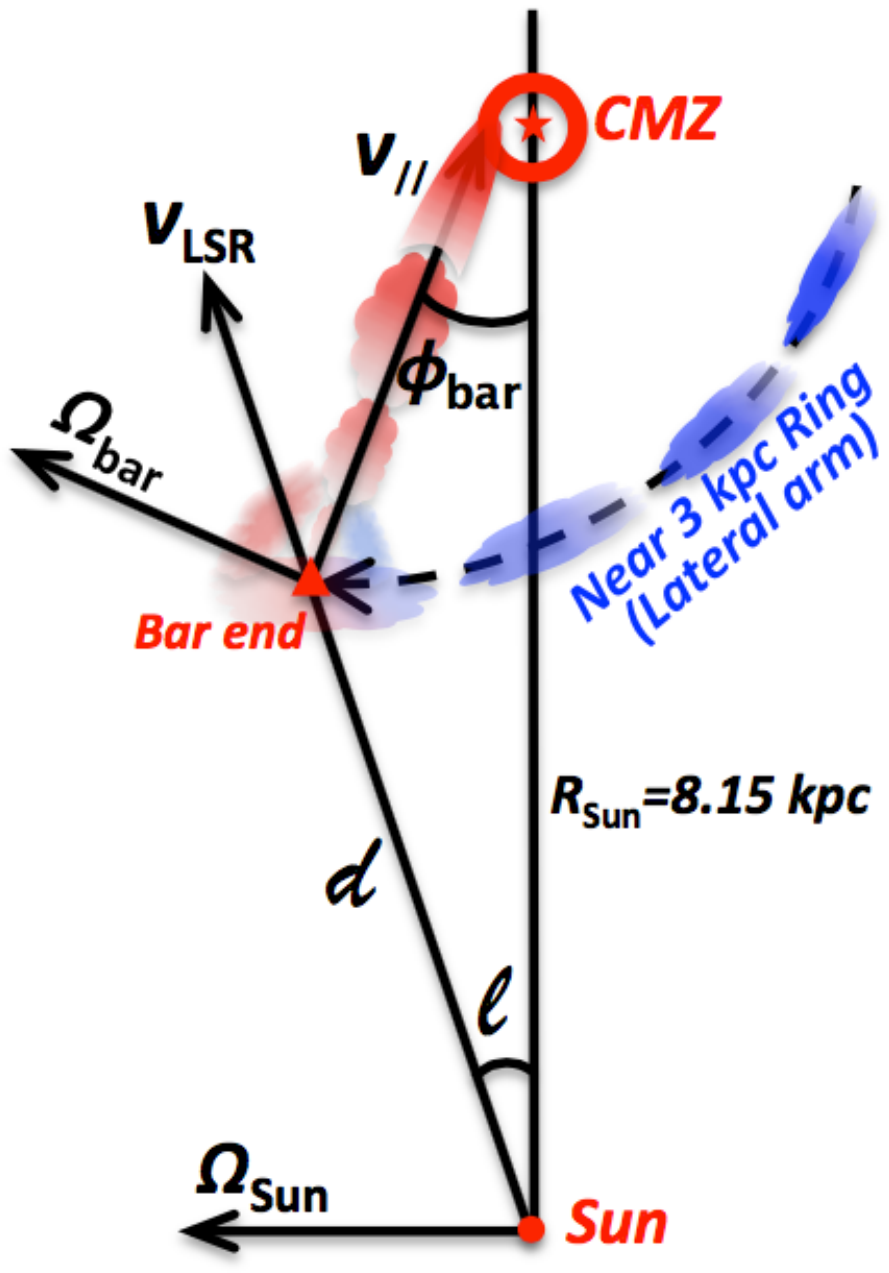}{0.6\textwidth}{} 
         }
\vspace{-19ex}
\caption{
Schematic representation of the WGL and the lateral arm
based on the morphological and kinematic features from the MWISP CO data.
In the diagram, the red shaded blocks represent the MCs flowing towards
the CMZ, while the blue shaded blocks show the MCs along the near 3~kpc
ring (or the lateral arm dragged by the rotation bar).
The end of the WGL intersects with the 3~kpc-ring structure at a distance 
of about 5~kpc. 
The bar plays an extremely important role in driving flows and transporting
the reservoir gas at $R_{\rm GC}\sim$3--4 kpc to the CMZ.
The red ring represents the CMZ, while the red dots, red stars, and red triangles 
represent the Sun, the GC, and the bar end, respectively.
The direction of the moving gas and the Sun is labeled by the black arrows.
Here, the longitude of the near bar end, $l$, is measured to be about $15^{\circ}$.
The inclination angle and the pattern speed of the bar are estimated to be 
$\phi_{\rm bar}=23^{\circ}\pm3^{\circ}$ and
$\Omega_{\rm bar}\lsim32.5\pm2.5 \km\ps\kpc^{-1}$ (see text), respectively.
\label{fig:cmz}}
\end{figure}

The features of the compressed and shocked molecular gas are well delineated
by the CO data, both along the body of the WGL and at the end of the WGL
(or the bar-spiral intersection region).
Figure~\ref{fig:cmz} shows a schematic view of the observation results.
In the diagram, we show that considerable gas is flowing towards the GC, 
forming the WGL connecting the 3~kpc-ring and the CMZ. Here, the gaseous disk at 
$R_{\rm GC}\sim$3~kpc and the area around it becomes the main gas reservoir for the 
large-scale accreting flows inwards to the CMZ. 

The morphological and kinematic CO features of the WGL 
can be approximately explained by the gas rotation 
at a certain pattern speed driven by the Galactic bar potential.
In this way, the moving gas driven by the rotation bar \citep[e.g.,][]
{1992MNRAS.259..345A,1999A&A...345..787F,2019MNRAS.484.1213S} can be roughly described by \\
\begin{equation}
v_{\scriptscriptstyle {//}}=\frac{v_{\rm {LOS}}-(\Omega_{\rm {bar}}-\Omega_{\rm {Sun}})\times{\rm {sin}}(l)}{{\rm {cos}}(\phi_{\rm {bar}}+l)}
\end{equation}

Here, $v_{\scriptscriptstyle {//}}$ is the gas velocity towards the CMZ region 
along the bar (i.e.,
$\vec{v}=\vec{v}_{\scriptscriptstyle {//}}+\vec{v}_{\rm {rot}}$ 
in the inertial frame of the Galaxy)
and $v_{\rm {LOS}}$ is the measured velocity of the moving gas along the line-of-sight
(LOS); 
$l$, $\phi_{\rm {bar}}$ are the longitude of the MC in the dust lane and the 
Sun-Galactic centre angle, respectively; and 
$\Omega_{\rm {bar}}$ and $\Omega_{\rm {Sun}}$ are rotation speed of the bar and 
the Sun, respectively 
\citep[e.g., $\Omega_{\rm {Sun}}=30.32\pm0.27 \km\ps\kpc^{-1}$ in][]{Reid19}.
We discuss these parameters in the following sections based on the
observed large-scale CO features associated with the WGL.

\subsubsection{Angle between LOS of the GC and Bar Major Axis, $\phi_{\rm bar}$}
The inclination angle of the bar ($\phi_{\rm bar}$) with respect to 
the Sun's position is critical in the analysis of gas dynamics and 
gas structure characteristics related to the WGL. 
We have shown that the WGL extends up to about $l=15^{\circ}$. 
Interestingly, we find that maser G015.66-00.49 at a parallax of 0.22$\pm0.029$~mas 
and $v_{\rm {LSR}}=-4\km\ps$ \citep[][]{Reid19} is associated with the 
MC G15.365-0.541 at $v_{\rm {LSR}}=-0.87\km\ps$.
The MC G15.365-0.541, together with the nearby CO gas, displays a filamentary 
structure extended along the Galactic plane from $l=14\fdg6$ to $l=15\fdg7$
(see the CO gas in the purple rectangle in Figure~\ref{fig:f2}c).
Assuming that the elongated structure at a distance of $\sim4.9$~kpc 
(i.e., between the median and the upper limit of the maser distance) 
is located close to the end of the WGL, the length of the WGL is estimated to be 
$\lsim3.6$~kpc, which is slightly larger than the radius of the 3~kpc-ring
\citep[e.g., the Galactocentric distance of the 3~kpc-ring, $R_{\rm ring}\sim$3.1~kpc, 
see details and Figures 2 and 3 in][]{2021ApJ...910..131S} but reasonable.
Indeed, the estimated length of the WGL here agrees well with the bar's length
of $\approx$3.5--3.6~kpc from very recent observations
\citep[e.g.,][]{2023MNRAS.520.4779L,2024MNRAS.528.3576V}.

The value of $\phi_{\rm bar}$ is thus $\gsim20^{\circ}$
by considering the WGL has the longest length of 3.6~kpc at $l\sim15^{\circ}$.
This result agrees with the CO observation that the WGL cannot 
continue to maintain a positive inflow velocity at the point of $l\sim15^{\circ}$
(i.e., the gas should have $v_{\rm {LSR}}\gsim 0 \km\ps$ at $l\sim15^{\circ}$ 
for $v_{\scriptscriptstyle {//}} \gsim 0 \km\ps$; see Section 3.2.3).
Given the deficiency of atomic and molecular gas within the 3~kpc-ring
\citep[e.g.,][]{1976ARA&A..14..275B,2001ApJ...547..792D,2022ApJ...930..112S}, 
on the other hand, it is reasonable to define a minimum length 
of the WGL of $\gsim$~3.1~kpc. Given the uncertainties of the length of 
$R_{\rm bar}\approx$3.1--3.6~kpc and the distance to the bar end of 
$\gsim4.9$--5.2~kpc, the estimation
of $\phi_{\rm bar}$ is in the range $20^{\circ}$--$26^{\circ}$.

The value of $\phi_{\rm bar}$ from our estimation is in good agreement with 
the previous estimates of $20^{\circ}$--$30^{\circ}$ 
\citep[e.g.,][]{2013MNRAS.435.1874W,2015MNRAS.450.4050W,2016ARA&A..54..529B,
2017MNRAS.471.4323S,2020RAA....20..159S}.
Below we use the value of $\phi_{\rm bar}=23^{\circ}\pm3^{\circ}$ 
as an input for the derivation of other parameters.

\subsubsection{Tilted Angle between the Bar and the $b=0^{\circ}$ Plane, $\theta_{\rm bar}$}
Additionally, since the latitude values of the bar traced by the WGL
can be well presented by the CO distribution, the tilt angle of the bar 
relative to the $b=0^{\circ}$ plane can be roughly estimated by
$\theta_{\rm bar}={\rm {tan}}^{-1} ({\rm {sin}}(\phi_{\rm bar}) {\rm {tan}} (b) / {\rm {sin}} (l))$.
The tilt angle is thus $\sim -2\fdg6\pm0\fdg3$ at $l\sim1\fdg7$, 
$\sim -4\fdg7\pm0\fdg6$ at $l\sim5\fdg2$, $\sim -2\fdg1\pm0\fdg3$ at $l\sim10\fdg7$, 
and $\sim -0\fdg8\pm0\fdg1$ at $l\sim15\fdg2$.
These results \citep[also refer to][]{2020MNRAS.499.4455T} indicate 
that the WGL is mainly located below 
the Galactic plane at a typical scale of 30--130~pc.
The typical values of the gas lane clouds deviated from the $b=0^{\circ}$ plane
are comparable to or $\sim$3.5 times larger than 
the standard deviation of the vertical distribution of the CO clouds
in the inner Galactic molecular disk \citep[e.g.,][]{2021ApJ...910..131S}.

The total gas mass of the WGL is obviously concentrated below the Galactic plane, 
which is the exact opposite of the far dust lane at $l\lsim-1^{\circ}$\citep[e.g.,][]
{2006A&A...447..533L,2008A&A...477L..21M,2019MNRAS.484.1213S,2021MNRAS.502.5896A}.
Naturally, the twisted $\infty$ gas feature in the CMZ \citep[][]{2011ApJ...735L..33M}
appears to be related to the tilted inflow structures, which may represent 
the transfer of angular momentum from the gas reservoir at kpc scales to 
the GC region at tens of pc scales in the accretion-driven torques.
Moreover, the asymmetric features of the FBs are probably related to 
tilted nuclear outflows, which are directly associated with the 
tilted gas inflows at various scales from $\sim$kpc to $\sim$pc.

\subsubsection{Pattern Speed of the Galactic Bar, $\Omega_{\rm bar}$}
As an essential parameter for characterizing the bar dynamics, the bar pattern speed,
$\Omega_{\rm bar}$, can be measured from observations.
We have shown that the molecular gas driven by the Galactic bar has an 
LOS velocity of $\sim 0$~$\km\ps$ at $l_{\rm max}\approx15^{\circ}$. 
This result can be used to constrain $\Omega_{\rm bar}$.
That is, the radial velocities of the gas should be somewhat greater than 
$\sim 0$~$\km\ps$, i.e., $v_{\scriptscriptstyle {//}} \gsim 0 \km\ps$ near 
the bar end position or the bar-spiral intersection region, 
considering that the material is flowing inward 
to the CMZ region along the bar channel.

We estimate that the $\Omega_{\rm bar}$ is less than 
$\sim 30-35 \km\ps\kpc^{-1}$,
taking into account possible fluctuations in the observed LOS velocity of 
$\sim0$--$10 \km\ps$ for MCs close to the region of the WGL end.
We find that the estimated $\Omega_{\rm bar}$ here agrees well with 
$35-40 \km\ps\kpc^{-1}$ from the recent studies
\citep[e.g.,][]{2019MNRAS.488.4552S,2020RAA....20..159S,2024MNRAS.531L..14L,
2024arXiv240606678Z}.
In particular, our estimation is exactly the same as 
$\Omega_{\rm bar}= 33.29 \pm 1.81 \km\ps\kpc^{-1}$ from the VVV Infrared 
Astrometric Catalogue and Gaia DR2 data \citep[][]{2022MNRAS.512.2171C}.
This may imply that the WGL is well coupled with the stellar bar,
and the results tend to favor the slow-bar scenario in the current period.
The corotation radius, $R_{\rm CR}$, can be approximated as 5.4--6.3 kpc
based on the slower-bar model \citep[e.g., see][]{2020MNRAS.497..933H}.

It should be noted that the movement and evolution of the Galactic bar are closely 
related to the gravitational potential distribution and the bar's dynamics.
At the same time, the bar interacts with the adjacent material, leading to 
pattern fluctuations between the bar itself and its surrounding constituents.
Due to the highly dynamical process, the pattern speed of the Galactic bar 
may vary over time in one or several rotation periods of the bar 
(e.g., $\sim$200~Myr to several Gyr).
Indeed, the current slowing rate of the bar has been well constrained recently
\citep[e.g., $-4.5\pm1.4\ \km\ps\kpc^{-1}$~Gyr$^{-1}$ in][]{2021MNRAS.500.4710C}.
Finally, other factors may affect our estimate of $\Omega_{\rm bar}$ 
--for example, some viscous or turbulent processes associated with the gas driven 
by the rotation bar. However, further in-depth analysis of these effects 
is beyond the scope of this Letter. We only use the estimated value of 
$\Omega_{\rm bar}\lsim32.5\pm2.5 \km\ps\kpc^{-1}$ for the following discussions.

\subsection{Mass Inflow Rate into the CMZ}
Having determined the parameters of $\phi_{\rm bar}$ and $\Omega_{\rm bar}$, 
we can infer the process of the gas transport towards the CMZ region along the bar. 
The cyan color and cyan contours in Figure~\ref{fig:f1} display 
the CO and \HI\ distribution of the WGL along the bar.
The dominant material in the WGL is molecular gas,
and its total mass 
can be determined from the \twCO\ emission by using the
CO-to-H$_2$ conversion factor method.
We find that the value of $X_{\rm CO}$ near the GC is probably very different 
for the MCs in other regions \citep[e.g.,][]{1998A&A...331..959D,2007A&A...467..611F,
2013ARA&A..51..207B,2015ARA&A..53..583H,2023ApJ...959...93G,2024PASJ..tmp...44K}.

We use \twCO\ and \thCO\ lines to derive $X_{\rm CO}$ for MCs of the WGL 
in the local thermal equilibrium (LTE) method 
\citep[see, e.g.,][]{2015ApJ...811..134S}.
The area we focus on is between the region near the 3~kpc-ring and the CMZ region, 
and the estimated $X_{\rm CO}$ is about $\sim1.0\pm0.4\E{20}$~cm$^{-2}$(K~km~s$^{-1})^{-1}$
from 17 largest MC samples (angular size $>250\ {\rm {arcmin}}^2$) in the WGL
assuming $N({{\rm {H}}_2})\approx7\pm3\E{5}\times N({^{13}{\rm {CO}}})$.
Here, we find that the conversion factor from $N({^{13}}{\rm {CO}})$ to 
$N({{\rm {H}}_2})$ is between the value at the $R_{\rm {GC}}\sim$3~kpc region
\citep[i.e., $\sim4\E{5}$ from][]{1982ApJ...262..590F,2005ApJ...634.1126M} 
and that at the CMZ region \citep[i.e., $\sim5-10\E{5}$ from][]
{1989ApJ...337..704L,2010A&A...523A..51R,2018A&A...613A..42R}. 

The typical value of the $X_{\rm CO}$ for the WGL is 
$1.0\pm0.4\E{20}$~cm$^{-2}$(K~km~s$^{-1})^{-1}$, which is very close to 
$\sim(0.5-1.0)\E{20}$(K~km~s$^{-1})^{-1}$ for the $|l|\lsim10^\circ$ region 
from other methods \citep[e.g., see details in][]{2014A&A...566A.120S}.
The lower $X_{\rm CO}$ here is consistent with the lower
$I(^{13}$CO)/$I(^{12}$CO) ratio for the inflowing gas presented 
in Section 3.1.
For the purpose of studying comparison with other work, we tentatively use
the updated value of $X_{\rm CO}=1.0\E{20}$~cm$^{-2}$(K~km~s$^{-1})^{-1}$
\citep[about 0.5 times the often used value; e.g., see][]
{2001ApJ...547..792D,2004A&A...422L..47S,2013ARA&A..51..207B,2024MNRAS.527.9290K}
to estimate the total molecular mass of the WGL.

The molecular gas mass in the WGL is thus 1.1$\pm0.4\times10^{7}\ \Msun$ 
and the velocity gradient along the bar is about 80--90~$\km\ps\kpc^{-1}$.
In total, the gas mass is about 1.3$\pm0.4\times10^{7}\ \Msun$
by counting the additional atomic gas of about 0.2$\times10^{7}\ \Msun$ for the 
\HI\ emission associated with the WGL (see the emission near the black ellipses
in the upper panel of Figure~\ref{fig:f1} and the cyan contours in the bottom 
panel of Figure~\ref{fig:f1}).
Interestingly, the molecular gas masses per unit length of the large-scale flows
($\sim3\times10^{6}\ \Msun\kpc^{-1}$) are roughly comparable to those of 
the 3-kpc arm by adopting the same X-factor value here \citep[see][]{2008Dame}.  
Moreover this feature can be interpreted by the abundant gas dragged by a rotating bar,
which is shown on the mirror situation for the near 3-kpc structure with enhanced CO
at positive longitude and the far 3-kpc CO at negative longitude
\citep[see Figure 14 in][]{2016ApJ...823...77R}.
This symmetrical gas structure is more common in
the barred galaxies \citep[e.g., see][]{2023A&A...676A.113S}.

Assuming that the far dust lane at $l\lsim-1^{\circ}$ is approximately 
symmetrical with the WGL at $l\gsim+1^{\circ}$, the average mass inflow rate is 
thus about 1.1$\pm0.3\ \Msun$~yr$^{-1}$ in a typical transport period of $\sim$24~Myr 
for gas from the bar end to the CMZ.
Here we do not consider the impact of the overshooting effect 
\citep[e.g.,][]{2019MNRAS.484.1213S} because we do not account for 
the amount of overshooting gas reaccreting into the near dust lane
from the far-side lane either 
\citep[i.e., the CO gas towards us with a relatively small LSR velocity 
and the positive longitude; also see][]{2023ApJ...959...93G}.
The overshot gas may actually reaccrete and flow into the CMZ within several Myr
\citep[][]{2021ApJ...922...79H}, which is about 1 order of magnitude smaller 
than the typical transport period.
Taking into account the compensation for the overshooting gas
back from the far dust lane \citep[e.g., refer to the extended trajectory of 
red lines of Figure 15 in][]{2022ApJ...939...58W}, our estimate does not,
in principle, significantly overestimate the overall inflow rate from the
gaseous disk to the region near the CMZ.

The overall distribution of the WGL shows that the current mass transport
is relatively continuous, suggesting that the gas in the CMZ
can accumulate rapidly over a period of several tens of Myr.
However, the gas inflow rate to the CMZ seems to be comparable to the outflow rate 
\citep[e.g., $\sim1.5\pm0.5\ \Msun$~yr$^{-1}$, see details in][]{2022ApJ...930..112S}
driven by the Galactic nuclear winds after applying the updated lower X-factor above.
The general equilibrium between the outflow and the inflow rate likely indicates
that the accumulation of gas in the CMZ may take a relatively longer time of
hundreds Myr to several Gyr.
Our results agree well with the discussion in \citet{2023ASPC..534...83H}.

Moreover, the gas distribution along the bar has obvious characteristics 
of heterogeneity and clumping, leading to large variations 
in the instantaneous inflow rate and subsequent the accumulation of 
a large amount of materials in a small region.
This process has a significant impact on the star formation 
process and star formation rate in the local region.
For example, \cite{2020ApJ...901...51A} suggested that 
Sgr E formed upstream in the far dust lane of the Galactic bar a few 
Myr ago, which may be related to the rapid accumulation of gas in a small
region and the successive emergence of star groups therein.
Generally, the gas cycle on the long timescale dominates the large-scale 
structural features of the Galaxy as a whole, while the episodic processes, 
such as the local individual characteristics of gas flows and 
the following stellar/AGN feedback, may significantly regulate 
the dynamics and spatial distribution of the gas at parsec to hundreds of parsecs scales.

\section{Summary}
By using the MWISP CO data in the region of $l=1\fdg2-19\fdg0$ and $|b|\lsim3\fdg0$,
we take a closer look at the gas flowing towards the Galactic CMZ. 
The new findings and results are summarized as follows.\\

1. The MCs with arc-like and ballistic structures are well delineated
along the near dust lane in the region $l=[1\fdg2, 10\fdg3]$ and
$b=[0\fdg0,-2\fdg1$]. 
The moving and shocked MCs towards the CMZ are further confirmed by 
the corresponding wide CO broadening and the large-scale velocity gradient, 
as well as the continuous gas structure assembled by a series of 
bow-like and bright-edged MCs heading towards the GC. \\

2. In addition to the well-known near dust lane at $l\lsim10^{\circ}$,
more filamentary MCs up to $l\sim15^{\circ}$ are found to be the extension
of the near dust lane structure in the $l$-$b$-$v$ space.
In $l\sim10^{\circ}$--$15^{\circ}$, these elongated MCs,
together with cometary-like and long-shell gas structures, also have
a relatively large CO line broadening or velocity gradient.
These features are very similar to those MCs in the near dust lane
at $l\lsim10^{\circ}$. \\ 

3. The enhanced turbulence of the gas, together with the certain
velocity gradient of the long structure, may originate from large-scale
continuous gas flows with the uniform pattern of motion.
In fact, the leading edge of the radial bar shock is well outlined by
these interesting MCs with sharp edge and/or head-to-tail ballistics
towards the GC, as well as many bullet-shaped MC structures embedded
in the shock front.\\

4. The large-scale CO gas forms the WGL structure that connects the CMZ 
and the 3~kpc-ring. The morphological and kinematic CO features of the WGL provide 
robust evidence that a very large amount of molecular gas is moving 
towards the CMZ from the 3~kpc-ring.
We propose that the large-scale gas structure well represents
radial flows towards the CMZ and the tilted gas flow with considerable
acceleration is driven by the Galactic bar. \\

5. The WGL intersected to the $R_{\rm GC}\sim$3.1--3.6~kpc region is
accreting the surrounding material therein and 
transporting gas towards the CMZ.  
The inclination angle and the rotation speed of the WGL are
$\phi_{\rm bar}=23^{\circ}\pm3^{\circ}$ and 
$\Omega_{\rm bar}\lsim32.5\pm2.5 \km\ps\kpc^{-1}$, respectively. 
The result indicates that the large-scale CO structure seems to be coupled 
with the stellar bar of our Milky Way.   
The WGL is below the Galactic plane with a typical scale of 30--130~pc. \\

6. The total gas mass of the WGL is 1.3$\pm0.4\times10^{7}\ \Msun$
by adopting the $X_{\rm CO}=1.0\pm0.4\E{20}$~cm$^{-2}$(K~km~s$^{-1})^{-1}$ calculated
from the MWISP \twCO\ and \thCO\ data and the additional \HI\ gas. 
The estimated lower $X_{\rm CO}$ agrees with the observational fact
of lower $I(^{13}$CO)/$I(^{12}$CO) ratio for the MCs in the WGL.
The velocity gradient along the bar is about 80--90~$\km\ps\kpc^{-1}$, indicating the
typical transport period of $\sim$24~Myr for the gas inflows from the bar end 
or the bar-spiral intersection region to the CMZ.  \\

7. The average mass inflow rate towards the Galactic CMZ is thus estimated to be about 
1.1$\pm0.3$~$\Msun$~yr$^{-1}$ by considering the roughly symmetrical inflow structure 
of the WGL at the positive longitude and the far dust lane at the negative longitude. 
If we use the updated lower X-factor above for the outflow gas, 
the value of the gas inflow rate seems to be roughly comparable to the
outflow's rate of $\sim1.5\pm0.5\ \Msun$~yr$^{-1}$ for the Galactic nuclear winds.   \\

\acknowledgments
This research made use of the data from the Milky Way Imaging Scroll Painting 
(MWISP) project, which is a multiline survey in \twCO/\thCO/C$^{18}$O along the 
northern Galactic plane with the PMO 13.7m telescope. We are grateful to all the members 
of the MWISP working group, particularly the staff members at the PMO 13.7m telescope, 
for their long-term support. MWISP was sponsored by National Key R\&D Program of 
China with grants 2023YFA1608000 \& 2017YFA0402701 and by CAS Key Research Program 
of Frontier Sciences with grant QYZDJ-SSW-SLH047.
We also acknowledge support from the National Natural Science Foundation of China 
through grants 12173090 and 12041305. 
We are very grateful to the reviewer for the helpful and constructive 
comments that have largely improved this work.

The work makes use of publicly released data from the HI4PI survey that combines 
the EBHIS in the Northern Hemisphere with the GASS in the Southern Hemisphere.
The Parkes Radio Telescope is part of the Australia Telescope National Facility,
which is funded by the Australian Government for operation as a National Facility 
managed by CSIRO. The EBHIS data are based on observations performed with the 
100 m telescope of the MPIfR at Effelsberg. EBHIS was funded by the Deutsche 
Forschungsgemeinschaft (DFG) under grants KE757/7-1 to 7-3.
This publication makes use of data products from the Wide-field Infrared Survey 
Explorer, which is a joint project of the University of California, Los Angeles, 
and the Jet Propulsion Laboratory/California Institute of Technology,
funded by the National Aeronautics and Space Administration.

\facility{PMO:DLH}

\bibliography{references}{}
\bibliographystyle{aasjournal}

\clearpage

\clearpage

\clearpage

\end{document}